\documentclass[useAMS,usenatbib]{mn2e}

\usepackage{rotating}
\usepackage{multirow}

\usepackage{epsfig}
\usepackage{lscape}
\usepackage{longtable}
\usepackage{subfigure}

\usepackage{graphicx}
\DeclareGraphicsExtensions{.ps}

\makeatletter
\let\jnl@style=\rm
\def\ref@jnl#1{{\jnl@style#1}}

\def\aj{\ref@jnl{AJ}}                   
\def\araa{\ref@jnl{ARA\&A}}             
\def\apj{\ref@jnl{ApJ}}                 
\def\apjl{\ref@jnl{ApJ}}                
\def\apjs{\ref@jnl{ApJS}}               
\def\ao{\ref@jnl{Appl.~Opt.}}           
\def\apss{\ref@jnl{Ap\&SS}}             
\def\aap{\ref@jnl{A\&A}}                
\def\aapr{\ref@jnl{A\&A~Rev.}}          
\def\aaps{\ref@jnl{A\&AS}}              
\def\azh{\ref@jnl{AZh}}                 
\def\baas{\ref@jnl{BAAS}}               
\def\cjaa{\ref@jnl{ChJAA}}		
\def\jrasc{\ref@jnl{JRASC}}             
\def\memras{\ref@jnl{MmRAS}}            
\def\mnras{\ref@jnl{MNRAS}}             
\def\nar{\ref@jnl{NewAR}}               
\def\na{\ref@jnl{NewA}}                 
\def\pra{\ref@jnl{Phys.~Rev.~A}}        
\def\prb{\ref@jnl{Phys.~Rev.~B}}        
\def\prc{\ref@jnl{Phys.~Rev.~C}}        
\def\prd{\ref@jnl{Phys.~Rev.~D}}        
\def\pre{\ref@jnl{Phys.~Rev.~E}}        
\def\prl{\ref@jnl{Phys.~Rev.~Lett.}}    
\def\pasp{\ref@jnl{PASP}}               
\def\pasj{\ref@jnl{PASJ}}               
\def\qjras{\ref@jnl{QJRAS}}             
\def\skytel{\ref@jnl{S\&T}}             
\def\solphys{\ref@jnl{Sol.~Phys.}}      
\def\sovast{\ref@jnl{Soviet~Ast.}}      
\def\ssr{\ref@jnl{Space~Sci.~Rev.}}     
\def\zap{\ref@jnl{ZAp}}                 
\def\nat{\ref@jnl{Nature}}              
\def\iaucirc{\ref@jnl{IAU~Circ.}}       
\def\aplett{\ref@jnl{Astrophys.~Lett.}} 
\def\apspr{\ref@jnl{Astrophys.~Space~Phys.~Res.}}
\def\bain{\ref@jnl{Bull.~Astron.~Inst.~Netherlands}}
\def\fcp{\ref@jnl{Fund.~Cosmic~Phys.}}  
\def\gca{\ref@jnl{Geochim.~Cosmochim.~Acta}}   
\def\grl{\ref@jnl{Geophys.~Res.~Lett.}} 
\def\jcp{\ref@jnl{J.~Chem.~Phys.}}      
\def\jgr{\ref@jnl{J.~Geophys.~Res.}}    
\def\jqsrt{\ref@jnl{J.~Quant.~Spec.~Radiat.~Transf.}}
\def\memsai{\ref@jnl{Mem.~Soc.~Astron.~Italiana}}
\def\nphysa{\ref@jnl{Nucl.~Phys.~A}}   
\def\physrep{\ref@jnl{Phys.~Rep.}}   
\def\physscr{\ref@jnl{Phys.~Scr}}   
\def\planss{\ref@jnl{Planet.~Space~Sci.}}   
\def\procspie{\ref@jnl{Proc.~SPIE}}   

\begin{document}

\title{The column density distribution of hard X-ray radio galaxies}
 
\author[Francesca Panessa]{F. Panessa$^1$\thanks{E-mail: francesca.panessa@iaps.inaf.it},
L. Bassani$^2$, R. Landi$^2$, A. Bazzano$^1$, D. Dallacasa$^{3,4}$, F. La Franca$^5$, 
\newauthor A. Malizia$^2$, T. Venturi$^4$, P. Ubertini$^1$ \\
$^1$ INAF - Istituto di Astrofisica e Planetologia Spaziali di Roma (IAPS-INAF), Via del Fosso del Cavaliere 100, 00133 Roma, Italy\\
$^2$ Istituto di Astrofisica Spaziale e Fisica Cosmica (IASF-INAF), Via P. Gobetti 101, 40129 Bologna, Italy \\
$^3$ Dipartimento di Fisica e Astronomia, Universit\`a di Bologna, via Ranzani, 1, 40127, Bologna, Italy \\
$^4$ Osservatorio di Radioastronomia (ORA-INAF), Via P. Gobetti 101, 40129 Bologna, Italy \\
$^5$ Dipartimento di Matematica e Fisica, Universit\`a degli Studi Roma Tre, via della Vasca Navale 84, 00146 Roma, Italy  \\ 
}
\date{}

\maketitle

\begin{abstract}
In order to investigate the role of absorption in AGN with jets, we have studied the column density distribution of a 
hard X-ray selected sample of radio galaxies, derived from the INTEGRAL/IBIS and \emph{Swift}/BAT
AGN catalogues ($\sim$ 7-10\% of the total AGN population). The 64 radio galaxies have a typical FRII radio morphology and 
are characterized by high 20-100 keV luminosities 
(from 10$^{42}$ to 10$^{46}$ erg/s) and high Eddington ratios (Log L$_{Bol}$/L$_{Edd}$ typically larger than $\sim$ 0.01). 
The observed fraction of absorbed AGN (N$_{H}$ $>$ 10$^{22}$ cm$^{-2}$) is around 40\% among the total sample,
and $\sim$ 75\% among type 2 AGN. The majority of obscured AGN are narrow line objects, while unobscured AGN are broad line objects, 
obeying to the zeroth-order predictions of unified models. A significant anti-correlation between the radio core dominance
parameter and the X-ray column density is found. The observed fraction of Compton thick AGN is $\sim$ 2-3\%, in comparison with the 5-7\% found in radio-quiet hard X-ray selected AGN.
We have estimated the absorption and Compton thick fractions in a hard X-ray sample containing both radio galaxies and non-radio galaxies
and therefore affected by the same selection biases. No statistical significant difference was found in the absorption properties of radio galaxies and 
non radio galaxies sample. In particular, the Compton thick objects are likely missing in both samples and the fraction of obscured radio galaxies appears to decrease 
with luminosity as observed in hard X-ray non-radio galaxies.
\\

\end{abstract}


\section{Introduction}

Unification models for Active Galactic Nuclei (AGN) hypothesize a common structure for the central engine, where the
different orientation of the line-of-sight determines a different classification of the source (Antonucci 1993, Urry \& Padovani 1995). 
Material accretes around a central super massive black hole in the form of a disk, often surrounded by a corona of hot electrons responsible
of the emission at high X-ray energies. At larger scales, a dusty torus obscures the inner region of the AGN, reducing the amount of observable X-ray radiation. Moreover, the torus obscures
the optical broad emission lines produced in the broad line region located within parsec scales from the nucleus, but it does not obscure the optical narrow emission lines produced at kpc
scales from the nucleus. According to this picture, type 1 AGN show broad and narrow lines
in their optical spectra as they are viewed face-on, while type 2 AGN only display narrow emission lines as they are viewed edge-on. This picture is confirmed 
by the presence of large amount of obscuration in the X-ray spectra of type 2 AGN and the little absorption in type 1 (e.g., Risaliti et al. 1999, Bassani et al. 1999).

The fraction of obscured AGN is important for the understanding of the nature of the circumn-nuclear medium,
but it is also fundamental to characterize the history of black hole accretion in the Universe (Alexander et al. 2005, Gilli et al. 2007). 
Multi-wavelength surveys have attempted to constrain the number of obscured (N$_{H}$ $>$ 10$^{22}$ cm$^{-2}$) and Compton thick AGN (CT, N$_{H}$ $>$ 10$^{24}$ cm$^{-2}$),
in the optical (e.g., Risaliti et al. 1999, Cappi et al. 2006, Panessa et al. 2006, Akylas \& Georgantopoulos 2009, Jia et al. 2013, Vignali et al. 2014 ), infrared (e.g., Fiore et al. 2009, Alexander et al. 2011, Brightman \& Nandra 2011)
and hard X-rays (e.g., Sazonov et al. 2008, Malizia et al. 2009, Burlon et al. 2011). 
Hard X-rays surveys should be able to detect obscured AGN with less biases, however, even at energies $>$ 10 keV, part of the intrinsic flux is reduced by obscuration to a level below 
the survey detection limit of current missions, leading to an observed fraction of CT AGN of a few percent (e.g., Ajello et al. 2008, Malizia et al. 2009, 2012, Burlon et al. 2011, 
Vasudevan et al. 2013),
when correcting for biases this fraction rises to around 20-30\%.

The fraction of obscured AGN appear to decrease with increasing observed X-ray luminosity in the local and distant Universe
(e.g., Ueda et al. 2003, La Franca et al. 2005, Sazonov et al. 2007, Burlon et al. 2011, Ueda et al. 2014, Aird et al. 2015, Buchner et al. 2015),
this effect is likely due to the decrease of the covering factor of the obscuring material with the luminosity,
however this result has been questioned by Sazonov et al. (2015). 
A similar dependency has been found also with redshift (the fraction of absorbed objects increases toward higher redshifts), although this result is more controversial
(e.g., La Franca et al. 2005, Treister \& Urry 2006, Ueda et al. 2014);
even at $z$ $>$ 2 larger absorption fractions are found in high luminosity AGN (Iwasawa et al. 2012).

In this work we investigate the role of absorption in radio galaxies. Does the presence of a jet influence the level of obscuration in AGN?
Radio galaxies launch relativistic jets on large
scales (from sub-kpc to Mpc) and are strong emitters in the
radio band. Fanaroff \& Riley (1974) classified radio galaxies into two
classes, based on their radio power and morphology: the low
luminosity radio galaxies (FRI) usually show a core, and twin jets
(on average symmetric in brightness) which lose collimation at
some distance from the core to form radio lobes; in the high
luminosity sources (FRII) the radio jets are faint, or invisible,
and culminate in high surface brightness regions (hot spots),
with the radio lobes being backflow emission. This morphological
classification is not sharp, and intermediate and/or hybrid
morphologies can be found (Capetti, Fanti \& Parma, 1995).

Radio galaxies have been also classified 
in low-excitation radio galaxies (LERGs) and high-excitation radio galaxies (HERGs) based on the optical emission lines from the nuclei  
(see Hine \& Longair 1979). Most FRI and a subset of low radio power FRII are LERGs and also a few HERGs show an FRI morphology.
It is believed that LERGs are inefficiently accreting objects and that their radio through X-ray nuclear emission is produced within their small-scale jet,
while the emission in HERGs is likely accretion dominated (Chiaberge, Capetti \& Celotti 2002, Hardcastle, Evans \& Croston 2006).
Moreover, HERGs show evidence for strong cosmic evolution at all radio luminosities, indicating a tendency of being located at larger distances, while LERGs are consistent with little or no
evolution. A possible scenario is that HERGs might be fueled at relatively high rates in
radiatively efficient standard accretion discs by cold gas, perhaps brought in through mergers and interactions, and with some of the
cold gas leading to associated star-formation (Best \& Heckman 2012).
So far, no clear dividing factor between the FRI and FRII sources is found even when combining radio
luminosity, accretion mode, large-scale environment and host galaxy luminosity (Gendre et al. 2013).

Early X-ray studies of radio galaxies revealed a large fraction of obscured nuclei, with narrow line radio galaxies showing column densities between 10$^{21}$ and 10$^{24}$ cm$^{-2}$, in agreement
with the unification predictions (Sambruna et al. 1999). An anti-correlation between the absorbing column density and the radio core dominance parameter is found, 
suggesting that the absorption increases as the jet orientation angle decreases, in agreement with the unified models (Grandi et al. 2006, Wilkes et al. 2013). 
Most of the X-ray studies on radio galaxies are based on individual sources and only few works are performed on statistically significant samples. 
One example is the X-ray study of 22 radio galaxies which shows that FRI typically have much lower intrinsic absorption than FRII (Evans et al. 2006). 
Similarly, LERGs are generally X-ray weak with little or no absorption (Hardcastle et al. 2006), suggesting that the torus itself is missing or receeding as a consequence of the inefficient accretion regime (Lawrence 1991).
A large fraction of obscured AGN is also found in a larger sample of $\sim$ 40 narrow line radio galaxies (Hardcastle et al. 2009), but no CT radio galaxy is found.
Recently, a study of 38 high redshift (1$<$ z $<$ 2) low frequency selected (178 MHz) 3CRR radio galaxies recovers a CT fraction of $\sim$ 20\%  (Wilkes et al. 2013), found by means of multi frequency diagnostics to unveil CT candidates otherwise missed by simple hardness ratio arguments. 
However, in the local Universe, only a handful of CT radio galaxies are found (e.g., Hardcastle, Evans \& Croston 2006, Eguchi et al. 2009, Guainazzi et al. 2006, Guainazzi et al. 2004).
This very low fraction of CT radio galaxies is puzzling and whether it is due to evolution or selection effect must be clarified.

\begin{figure*}
\begin{center} 
\parbox{14cm}{
\includegraphics[width=0.4\textwidth,height=0.3\textheight,angle=0]{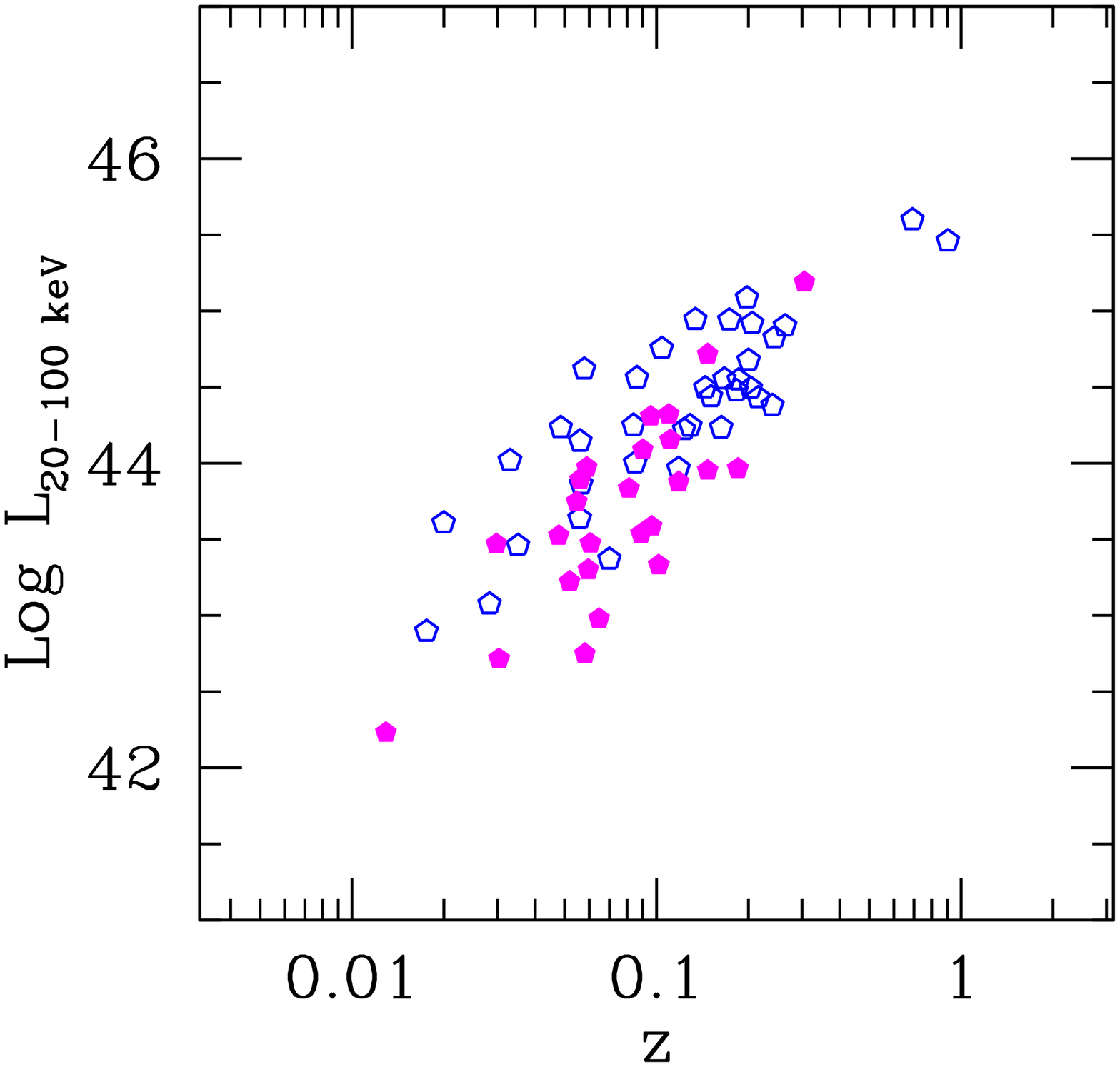}
\includegraphics[width=0.4\textwidth,height=0.3\textheight,angle=0]{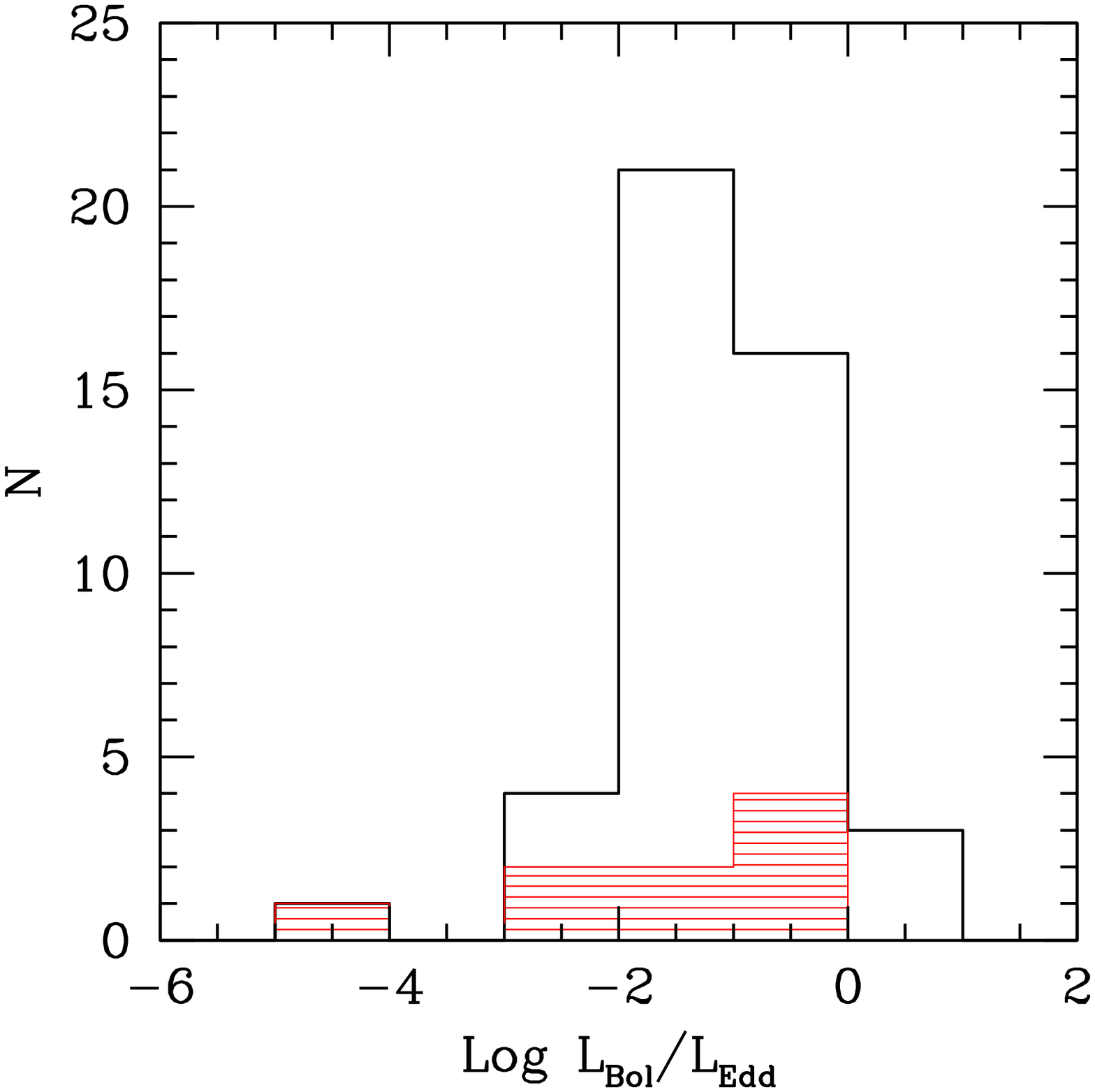}
}
\caption{Bilogarithmic plot of the 20-100 keV luminosity versus redshift (left panel), blue empty points are type 1 AGN, while magenta filled points are type 2 AGN. In the right panel, the distribution of the ratio between the bolometric 
luminosity and the Eddington luminosity for the total sample, where red shaded histogram represents FRI and intermediate FRI/FRII radio galaxies.}
\label{z}
\end{center}
\end{figure*}

Since 2002, INTEGRAL/IBIS and later on \emph{Swift}/BAT are surveying the sky at energies greater than 10 keV, releasing
all sky catalogues that contain a large number of AGN, some previously known and others discovered for the first time at hard X-rays (Bird et al. 2010; Krivonos et al. 2010; Cusumano et al. 2010; Baumgartner et al. 2013).
The hard X-ray selection is sensitive to gas rather than dust obscuration and should therefore allow the detection of highly obscured AGN, mostly mildly CT AGN with column densities in the range between
10$^{24}$ - 10$^{25}$ cm$^{-2}$ where the continuum radiation should be visible above 10 keV. 

In this work we present for the first time the column density distribution of a sample of radio galaxies selected at hard X-rays from INTEGRAL and \emph{Swift} observations.
The paper is organized as follows: in section 2 we present the radio galaxy sample and the X-ray data; in section 3 we report on the column density distribution and comment on some peculiar objects; in section 4 we discuss the fraction of obscured and CT AGN and in section 5 we present our conclusions.
Throughout this paper we assume a flat $\Lambda$CDM cosmology with ($\Omega_{\rm M}$,  
$\Omega_{\rm\Lambda}) = (0.3$,0.7) and a Hubble constant of 70 km s$^{-1}$ Mpc$^{-1}$ (Jarosik et al. 2011). 

\section{The Sample and the X-ray data}

For the purpose of this work we have used the same sample of radio galaxies as in Bassani et al. (2016, accepted), 
except for three candidate sources for which the radio galaxy classification is uncertain and that we have not considered in this work. 
In Bassani et al. (2016), AGN from hard X-ray catalogues have been selected looking for extended radio emission. The hard X-rays catalogues considered are: 
(1) the sample of 272 INTEGRAL AGN discussed by Malizia et al. (2012), with a few 
additional sources discovered or identified as AGN afterwards (Landi et al. 2010, Masetti et al. 2013, Krivonos et al. 2012), and 
(2) the 70 months {\it Swift}/BAT AGN catalogue of Baumgartner et al. (2013) which lists 822 objects associated with AGN or galaxies; 
a sample of 65 objects classified as 'unknown' has also been included in an attempt to uncover all possible radio galaxies in the BAT sample. 
In Bassani et al. (2016), the NRAO VLA Sky Survey (NVSS, Condon et al. 1998) and the Sydney University Molonglo Sky Survey (SUMSS, Mauch et al. 2003) 
were used to assess the morphology and derive the radio power of the hard X-ray selected AGN.
The morphological classification of our sample follows
Fanaroff \& Riley (1974), and is based on the radio morphology
(core, jets, lobes and hotspots) and radio power; the transition
power between FRI and FRII being roughly set at log P(1.4 GHz) $=$
24.5 W/Hz (Owen \& Laing 1989).

The final sample is made of 64 hard X-ray selected radio galaxies, 22 from Malizia et al. (2012) plus four additional INTEGRAL 
sources lately identified as radio galaxies\footnote{IGR J13107-5626 from Bassani et al. (2016),
IGR J14488-4008 from Molina et al. (2015), IGR J17488-2338 from Molina et al. (2014) and 4C+21.55 from Bassani et al. (2016).
and 60 from the {\it Swift}/BAT AGN catalogue. 
Twenty-two sources have been detected both by INTEGRAL and {\it Swift}/BAT.
Radio galaxies represent $\sim$ 7-10\% of the total AGN population at hard X-rays.

The sample is presented in Table~\ref{tab1}. In column three we report the optical classification of the sample (see Bassani et al. 2016)}:
22 AGN are of type 1, 19 of type 2 and 21 of intermediate class. The intermediate AGN are themselves split in 12 objects 
of early type (1.2-1-5) and 9 of late type (1.8-19) Seyfert classification. One source in the sample, IGR J13107-5626, has no redshift nor an optical classification, while PKS1737-60 has a photometric redshift and no optical classification.
The majority of the sample has a radio morphological classification as FRII, with only six FRI and six intermediate FRI/FRII. 
FRII are likely associated to high excitation radio galaxies (Buttiglione et al. 2009) accreting at high Eddington ratios, in this sense, their dominance is expected in a hard X-ray selected sample.

The X-ray fluxes and column densities are taken from the literature (see column 9 in Table~\ref{tab1} for references) except for 16 objects for which we have analyzed XRT+BAT broad band spectra (see Appendix A for details on the X-ray analysis).
In Figure~\ref{z} we plot the 20-100 keV X-ray luminosity versus redshift (left panel). It is clear that hard X-rays selects nearby very luminous radio galaxies, with 20-100 keV luminosities ranging from 10$^{42}$ to 10$^{46}$ erg/s. 
Black hole masses have been collected from the literature (column 11 for references), ranging from $\sim$10$^{7}$ to 10$^{9}$ M$_{\odot}$\footnote{Different methods have been used in literature to estimate the black hole mass, therefore the values reported
are not homogeneously derived, introducing a source of uncertainty which is difficult to estimate.}.
In Figure~\ref{z} (right panel) we plot the distribution of the ratio between the bolometric luminosity and the Eddington luminosity
(the bolometric luminosity is derived by applying a correction factor of 20 to the 2-10 keV luminosity, see Vasudevan \& Fabian 2007).
The Eddington ratios are relatively high suggesting that the hard X-ray radio galaxies are efficiently accreting sources and that their X-ray emission 
is likely dominated by the disk-corona system even in those galaxies with the most powerful jets (e.g., Reynolds et al. 2015).
The only source accreting at a low Eddington rate is Cen A, for this source it has been proposed that 
the hard X-ray emission is dominated by an Advection Dominated Accretion Flow (e.g., Fuerst et al. 2015)
and its detection is likely due to its closeness.

\begin{figure*}
\begin{center}
\parbox{14cm}{
\includegraphics[width=0.4\textwidth,height=0.3\textheight,angle=0]{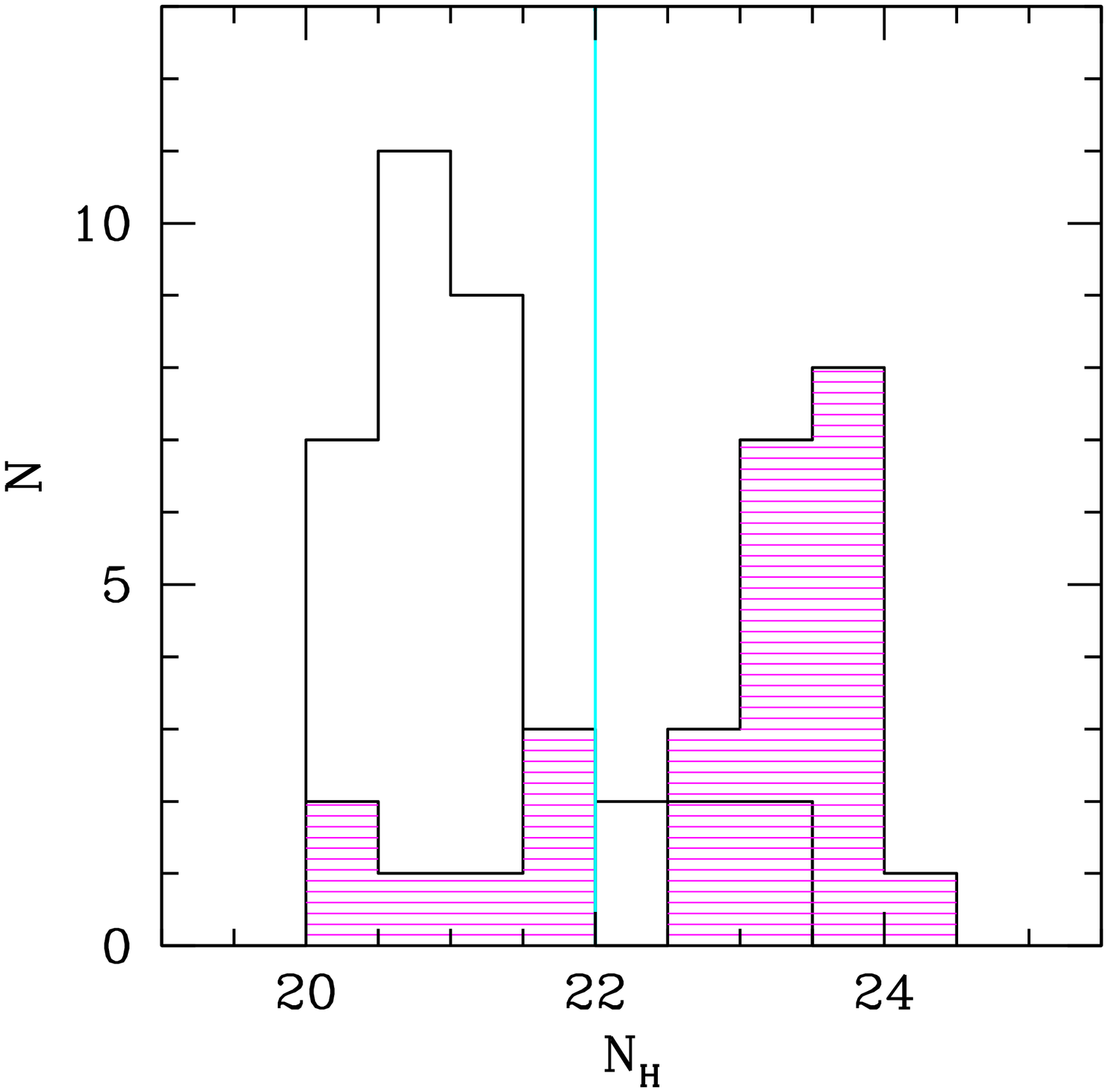}
\includegraphics[width=0.4\textwidth,height=0.3\textheight,angle=0]{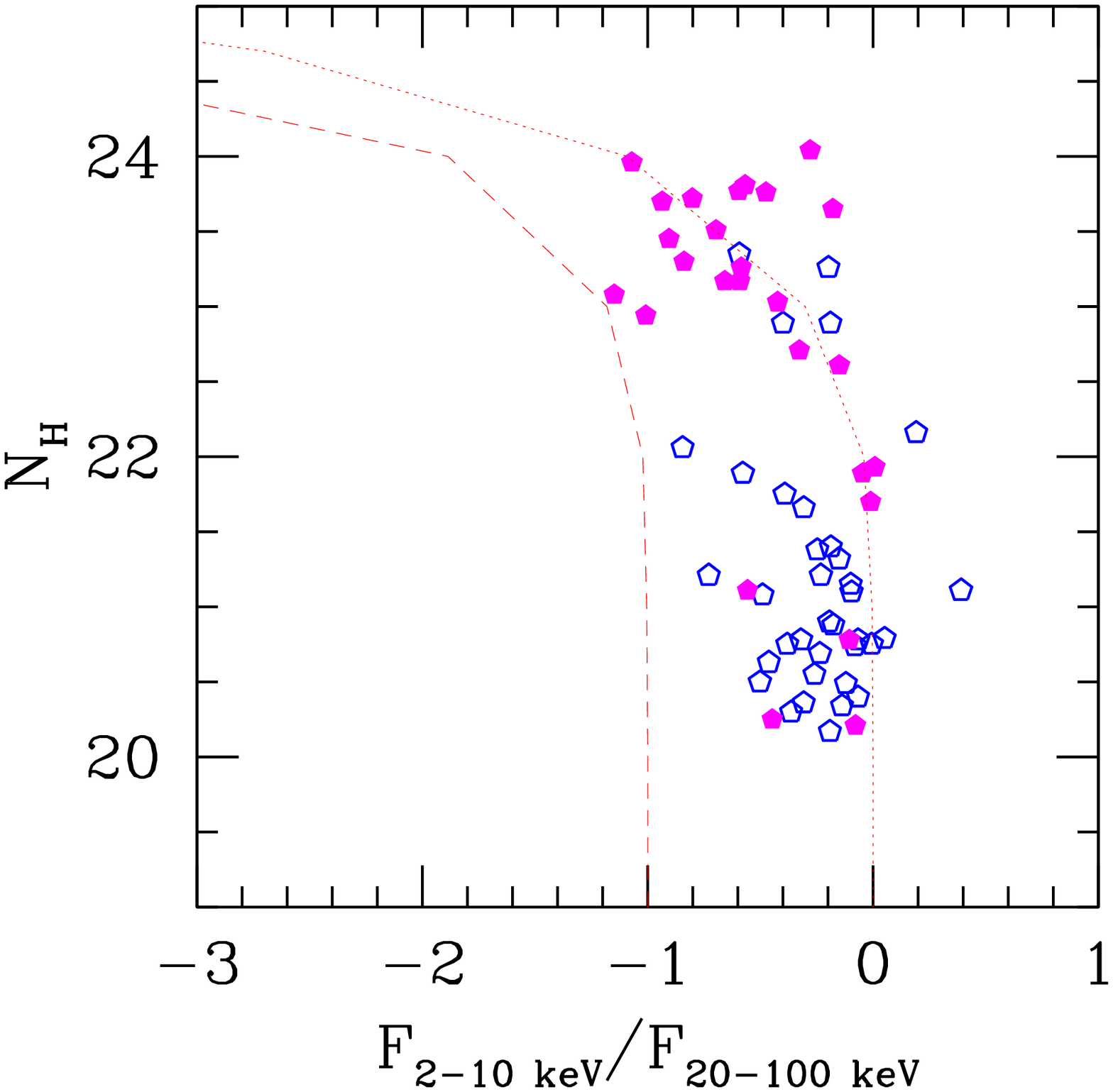}
}
\caption{Column density distribution in cm$^{-2}$, type 1 are represented by the black histogram while type 2 by the magenta shaded histogram (left panel). A cyan line divides 
obscured objects (N$_{H}$ $>$ 10$^{22}$ cm$^{-2}$) from unobscured objects. 
Column density distribution in cm$^{-2}$ versus the F$_{2-10 keV}$/F$_{20-100 keV}$ (right panel). Blue empty points are type 1 AGN, while magenta filled points are type 2 AGN.
Lines correspond to expected values for an absorbed power law with photon indices 1.5 (dot) and 1.9 (dash) as in Malizia et al. (2012).}
\label{nh}
\end{center}
\end{figure*}

\section{The column density distribution}

The column density distribution of radio galaxies has been studied in samples selected either in the X-rays (Sambruna et al. 1999) or in the radio (Evans et al. 2006, Hardcastle et al. 2009, Wilkes et al. 2013).
Narrow line radio galaxies typically show large X-ray column densities in agreement with the unified models, however the number of CT AGN has always been very small, quite below the 
20-50 \% fraction observed in local non-radio galaxies samples.
In the left panel of Figure~\ref{nh} we report the column density distribution of type 1 (black histogram) and type 2 AGN (blue shaded histogram). 
Intermediate Seyfert galaxies of early type classification (1.2-1.5) have been included in the type 1 group, while the late type (1.8-1.9) intermediates are included in the type 2 group.
As expected, type 2 radio galaxies are typically more absorbed than type 1, with the distribution peaking in the range between 10$^{23}$ cm$^{-2}$ and 10$^{24}$ cm$^{-2}$. 
In particular sources with N$_{H}$ $>$ 10$^{22}$ cm$^{-2}$ are 26/64 ($\sim$ 40$^{+13}_{-11}$ \%) among the total sample and 21/28 (75$^{+13}_{-9}$ \%) among type 2 Seyferts in agreement with the observed fractions found in optically selected samples of non-radio galaxies.
On the other hand only one source is CT (NGC 612) while VII Zw 292 has a column density consistent with the CT limit within an error of 90\%,
implying that the observed fraction of CT AGN is around 2-3\% ($<$ 11\%) among the total sample and 4-7\% ($<$ 23\%) among type 2 sources\footnote{Throughout 
the paper the standard error on the sample proportions in the binomial standard deviation for small samples are calculated at 2 $\sigma$ (Gehrels 1986).}. 

When low quality X-ray spectra do not allow a proper estimate of the real column density, it is possible to miss some CT AGN in a given sample. A diagram proposed by Malizia et al. (2007) uses the N$_{H}$ versus softness ratio (F$_{2-10 keV}$/F$_{20-100 keV}$) to unveil hidden CT AGN, its validity has been proven by several other works (Ueda et al. 2007, Winter et al. 2008, Matt et al. 2012). Misclassified CT objects populate the part of the diagram with low absorption and low softness ratios, as the 2-10 keV flux is suppressed by absorption, while the 20-100 keV is less affected by absorption.
The N$_{H}$ versus the ratio between the observed 2-10 keV and 20-100 keV of the radio galaxy sample is plotted in Figure ~\ref{nh} (right panel): no source populates the bottom left part of the diagram suggesting that there are no hidden mildly CT AGN in the sample.

\subsection{X-ray absorption versus optical and radio classification}

Similarly to what found in non-radio galaxies hard X-ray samples, a small fraction of objects are 'outliers', i.e., exceptions to the unified models such as type 1 AGN with absorption and type 2 AGN without absorption (see Malizia et al. 2012).
Six objects with broad emission lines (type 1, type 1.2 and type 1.5) display column densities $>$ 10$^{22}$ cm$^{-2}$. IGR J17488--2338 and 4C50.55 are absorbed by one (Malizia et al. 2014) and two (Molina et al. 2007) layers
of cold material partially covering the central source, respectively. Similarly in 3C 227 and PKS 2135-14 intrinsic absorption has been measured from the Chandra spectrum (Mingo et al. 2014).
The XMM-Newton spectrum of IGR J14488-4008 is characterized by absorption due to ionized elements (Molina et al. 2015) and the presence of a warm absorber component is also found in 3C445 (Torresi et al. 2012).
It is therefore likely that in most of these type 1 AGN ionized material located just off the torus and/or accretion disc might be responsible for the X-ray absorption.

Type 2 objects without X-ray absorption are another type of 'outliers' with respect to the unified models, where a column density $<$ 10$^{21}$ cm$^{-2}$ is not enough to obscure the BLR and the lack of broad emission lines in the optical spectrum is likely due to an absent or very weak broad line region (Panessa \& Bassani 2002, Bianchi et al. 2012). In this sample, only PKS 2331--240 is a type 2 radio galaxy without absorption and it is the 
subject of a recent multi-wavelength observational campaign conducted by our team, which will be presented in a forthcoming publication.

Osterbrock (1981) introduced the notations Seyfert 1.5, 1.8 and 1.9, where the subclasses are based on the optical appearance of the spectrum, with the numerically larger subclasses having weaker broad-line components relative to the narrow line ones.
According to unified models, at these intermediate classifications should correspond an intermediate inclination of the line of sight with respect to the obscuring torus, therefore type 1.8-1.9 should be mildly absorbed 
at X-rays (Risaliti et al. 1999). In our sample three sources are classified as type 1.8 Seyferts (3C 059, B3 0749+460A and PKS 1916-300) and show absorption consistent with the Galactic one,
suggesting that the broad line components maybe intrinsically weak. However, Trippe et al. (2010) have shown that a large fraction of 1.8 and 1.9 were misclassified due to an 
overestimation of the flux of broad H$_{\alpha}$, while others received this designation because of their low continuum flux. Accurate observations at multi epochs are therefore necessary to explore these possibilities.

In previous surveys, FRI radio galaxies have usually shown less absorption than FRII. In this sample, FRI and intermediate FIR/FRII, though numerically inferior, display a column density distribution similar to that of FRII. 
Interestingly, two out of five of the FRI in the sample are heavily absorbed 
(Cen A and PKS 2014-55). Three out of six intermediate FRI/FRII show N$_{H}$ $>$ 5 $\times$ 10$^{22}$ cm$^{-2}$, i.e., 4C+29.30, 3C433 and NGC~612. 

\subsection{The radio core dominance versus absorption}

\begin{figure}
\centering
\includegraphics[width=0.8\linewidth]{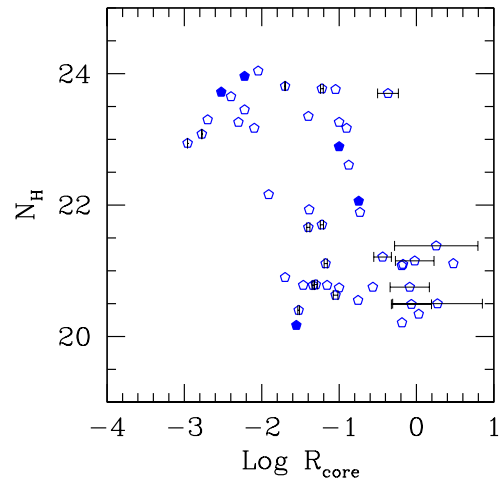}
\caption{Column density versus the radio core dominance parameter R$_{core}$ estimated at 5 GHz. Filled points have a R$_{core}$ estimated at 1.4 GHz.}
\label{corenh}
\end{figure}

The radio core dominance is defined as R$_{core}$ $=$ S$_{core}$/(S$_{tot}$ - S$_{core}$), where S$_{core}$ and S$_{tot}$     
are the core and the total flux densities at 5 GHz, respectively. It has been used as an orientation indicator,
as objects with a large R$_{core}$ should emit their radiation in a direction closer to the line of sight.
Indeed, a strong anti-correlation between the X-ray column density and R$_{core}$ has been found in samples of radio galaxies (e.g., Grandi et al. 2006, Wilkes et al. 2013).
A significant anti-correlation between R$_{core}$  and the inclination of the dusty torus (derived from modeling the infrared spectral energy distribution) has
also been found (Drouart et al. 2012).
We have collected the R$_{core}$ values at 5 GHz~\footnote{For five objects data were available at 1.4 GHz, while
R$_{core}$ in Fan \& Zhang (2003) were calculated using extended luminosities at 1.4 GHz, assuming two possible 
values of the spectral index ($\alpha$ $=$ 0.5 and $\alpha$ $=$ 1.0) to convert
the 1.4 GHz luminosity into 5 GHz luminosity (see their work for details).}.
We have correlated the X-ray column density with R$_{core}$ (see Figure~\ref{corenh}). The Spearman's Rho correlation coefficient is -0.50
with a two-tailed value of the null hypothesis probability of correlation of 0.00027, suggesting that also among our sample
of radio galaxies the two quantities are anti-correlated, in agreement with unified models.
No correlation is found between R$_{core}$ and the 2-10 keV luminosity
nor the 20-100 keV luminosity.
Note, however, that the calculation of R$_{core}$ parameter relies on data taken
at different epochs, with different instruments and at different linear scales. An appropriate set of radio images at the same frequency and with about
the same linear resolution is fundamental in this respect for a proper determination of the radio core dominance parameter.

\begin{figure}
\centering
\includegraphics[width=0.8\linewidth]{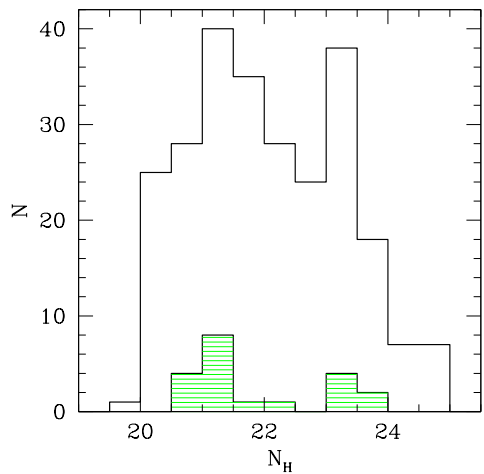}
\caption{Column density distribution of non-radio galaxies (black histogram) and radio galaxies (shaded green) of Malizia et al. (2012).}
\label{nhmal}
\end{figure}

\begin{figure*}
\begin{center}
\parbox{14cm}{
\includegraphics[width=0.4\textwidth,height=0.3\textheight,angle=0]{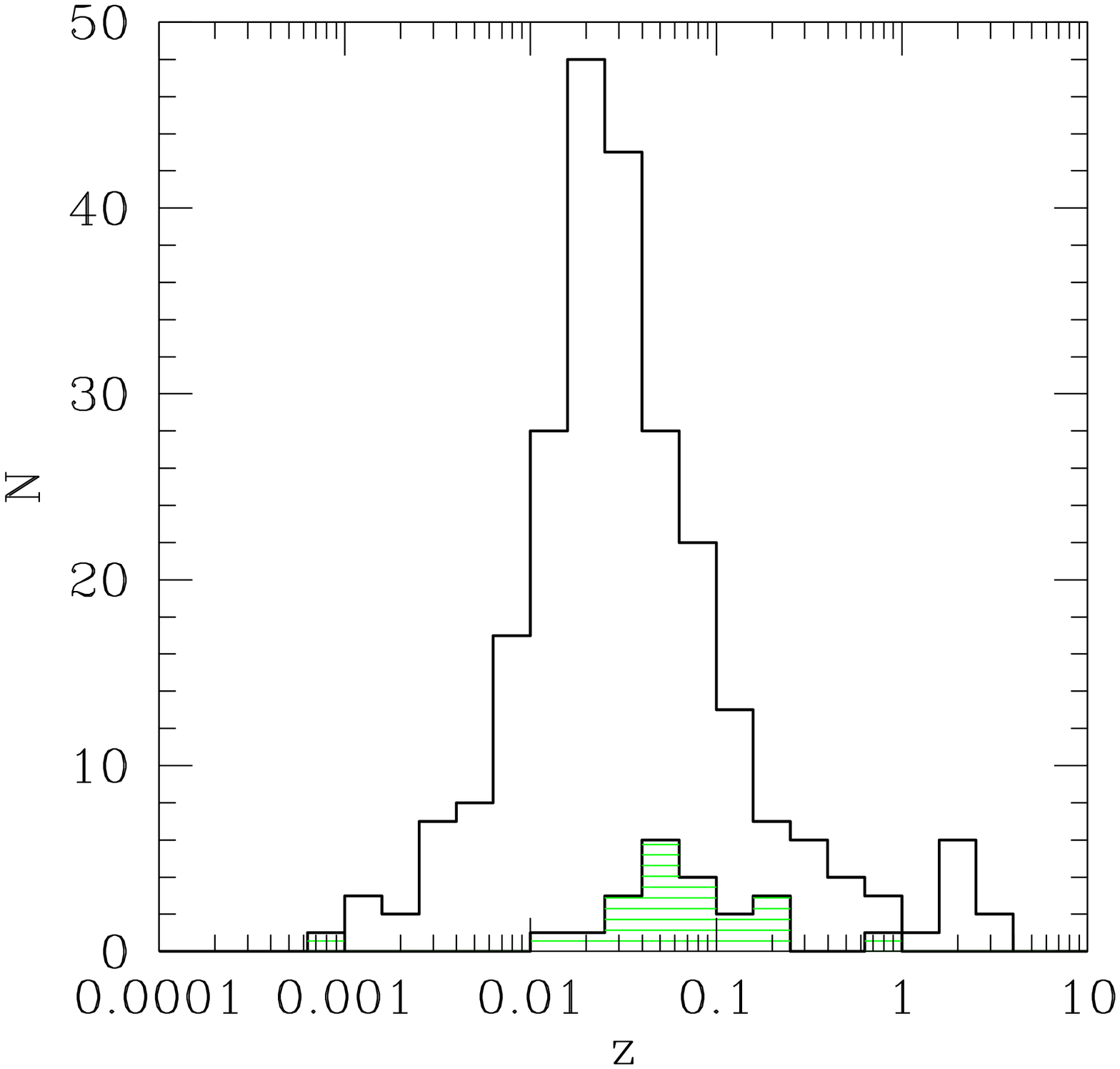}
\includegraphics[width=0.4\textwidth,height=0.3\textheight,angle=0]{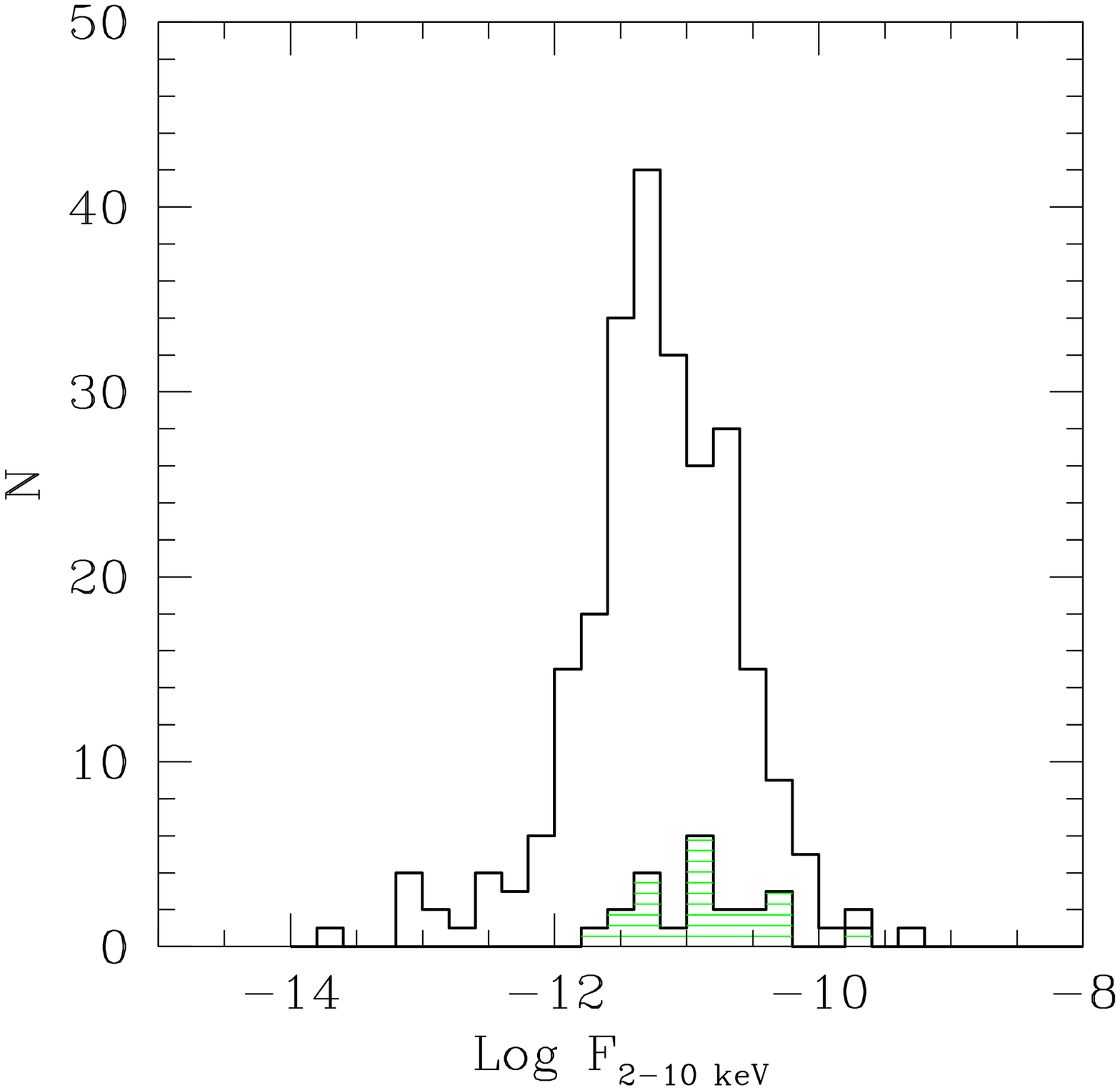}
}
\caption{Left panel: redshift distribution of non-radio galaxies and radio galaxies (shaded green) of the Malizia et al. (2012) INTEGRAL sample. Right panel:  2-10 keV flux distribution of non-radio galaxies and radio galaxies (shaded green) of the Malizia et al. (2012) INTEGRAL sample.}
\label{fff}
\end{center}
\end{figure*}

\section{The fraction of absorbed and Compton thick radio galaxies}

The determination of an intrinsic column density distribution, not affected by biases, is a hard task, even for the current hard X-ray surveys, 
where above 10-15 keV a significant fraction (up to 50-60\%) is missed for N$_{H}$ $>$ 3 $\times$ 10$^{24}$ cm$^{-2}$ (Malizia et al. 2009, Ajello et al. 2009).
Therefore the absorption bias limits the detection of Compton-thick objects only to those with bright fluxes and in the very local Universe.
Malizia et al. (2009) recover a fraction of 20-30\% of CT by limiting the sample to z$<$0.015, similarly Burlon et al. (2011) have found a fraction of 20\% CT AGN
by correcting for the absorption bias in the {\it Swift}/BAT sample.
In order to evaluate the distance and flux biases among hard X-ray selected radio galaxies, we have estimated the fraction of absorbed and CT AGN 
in a sample containing both non-radio galaxies and radio galaxies selected at hard X-rays and therefore, subject to the same biases effects. 
We have considered the total INTEGRAL AGN sample in Malizia et al. (2012) in which 22 out of 271 AGN are radio galaxies.
In Figure~\ref{nhmal} we present the column density distribution of non-radio galaxies (black histogram) and radio galaxies (green shaded histogram).
The fraction of absorbed AGN is 49$^{+6}_{-7}$\% (121/249) among non-radio galaxies compared to 36$^{+21}_{-16}$\% (8/22) among radio galaxies, while the fraction of CT AGN is 6$\pm$3\%
(14/249) compared to 0/22 ($<$1. 3\%) among radio galaxies.

In Figure~\ref{fff} (right panel), we have compared the redshift distribution of non-radio galaxies and radio galaxies (in green) separately. 
We have tested the null hypothesis that the two samples are taken
from populations with the same redshift distribution using the Kolmogorov--Smirnov (K-S) test at a significance level of $\alpha$ = 0.01. 
The derived p-value is 0.0014, therefore the null hypothesis can 
be rejected at 1\% level.
It is remarkable that no radio galaxies are found below z$<$0.01 (except for the peculiar source Cen~A)
in agreement with the fact that HERGs are preferentially found at larger distances than LERGs.

Excluding Cen~A, we have selected all non-radio galaxies in Malizia et al. (2012) in the redshift regime where all radio galaxies are found, i.e. z$>$0.01.
We have then compared the column density distribution of the two selected sub-samples.
At these redshifts, the observed fraction of absorbed AGN among non-radio galaxies is 42$^{+7}_{-6}$\% (89/212) consistent with the observed fraction
in radio galaxies of 33$^{+22}_{-16}$\% (7/21), similarly
the fraction of CT sources among non-radio galaxies is 9/212 (4$^{+4}_{-2}$\%)
compared with that of radio galaxies (0/21, $<$1. 4\%). 
In Figure~\ref{fff} (left panel), the 2-10 keV flux distribution of non-radio galaxies and radio galaxies (in green) is shown. In this case,
the derived KS p-value is 0.047, therefore the null hypothesis that the two samples are taken
from populations with the same flux distribution cannot be rejected at 1\% level.
No radio galaxies are found below $\sim$ 10$^{-12}$ ergs cm$^{-2}$ s$^{-1}$, therefore we have selected 
all non-radio galaxies with F$_{2-10 keV}$ $>$ 10$^{-12}$ ergs cm$^{-2}$ s$^{-1}$.
In this case the fraction of CT objects among non-radio galaxies (14/227, 6$^{+4}_{-2}$\%)
is only slightly larger than that of radio galaxies, 0/22 ($<$1. 3\%), while the fraction of absorbed AGN in non-radio galaxies (107/227, 47$^{+7}_{-6}$\%) 
is again consistent within errors with that of radio  galaxies (8/22, 36$^{+21}_{-16}$\%).
The above results suggest that radio galaxies might be affected by the same redshift bias 
as non-radio galaxies, suggesting that we are probably missing CT AGN in both classes of objects. However, the marginal evidence of a lower fraction of CT AGN in radio galaxies 
deserves deeper investigation through larger samples of objects.

\begin{figure*}
\begin{center}
\parbox{14cm}{
\includegraphics[width=0.4\textwidth,height=0.3\textheight,angle=0]{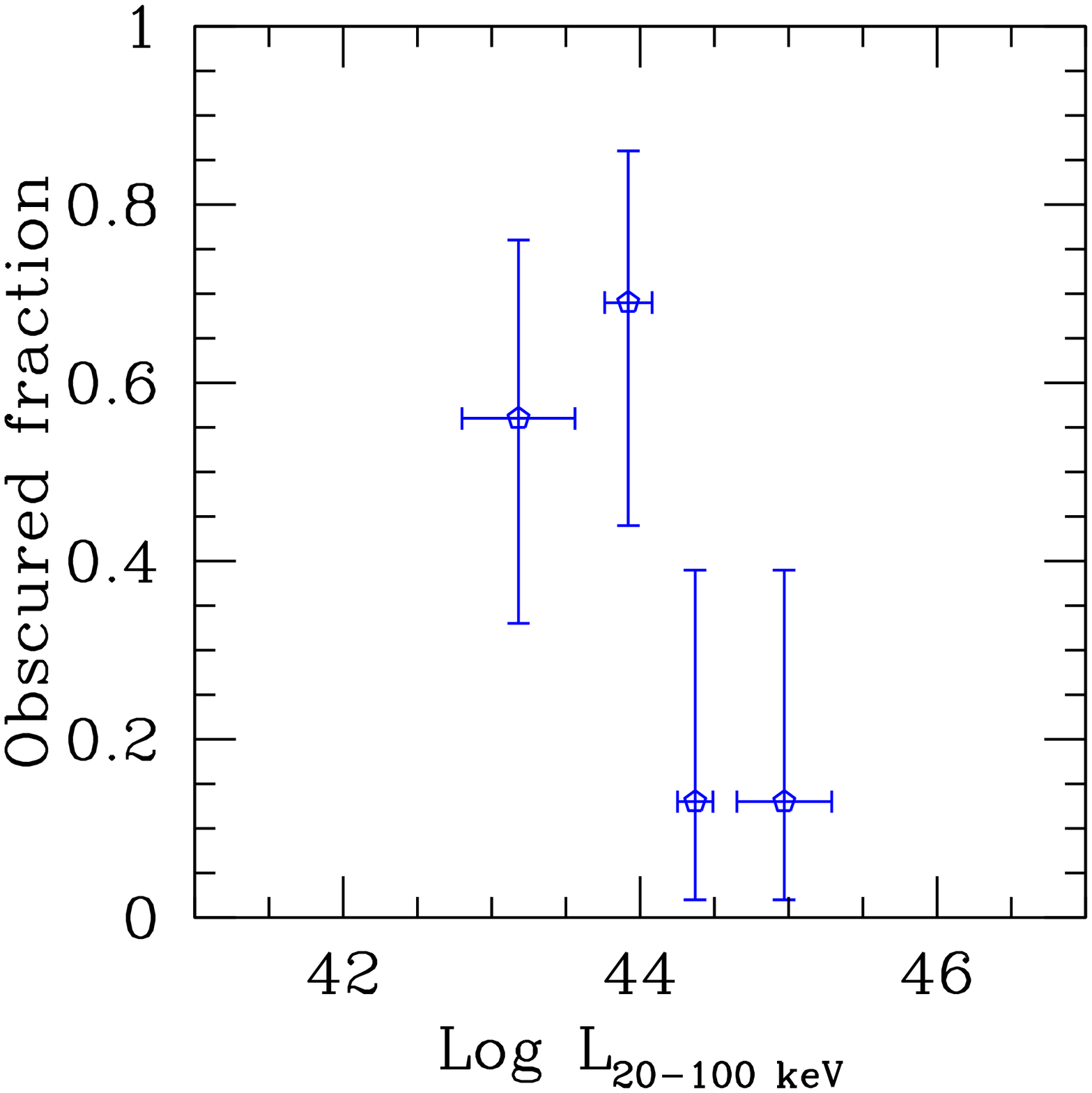}
\includegraphics[width=0.4\textwidth,height=0.3\textheight,angle=0]{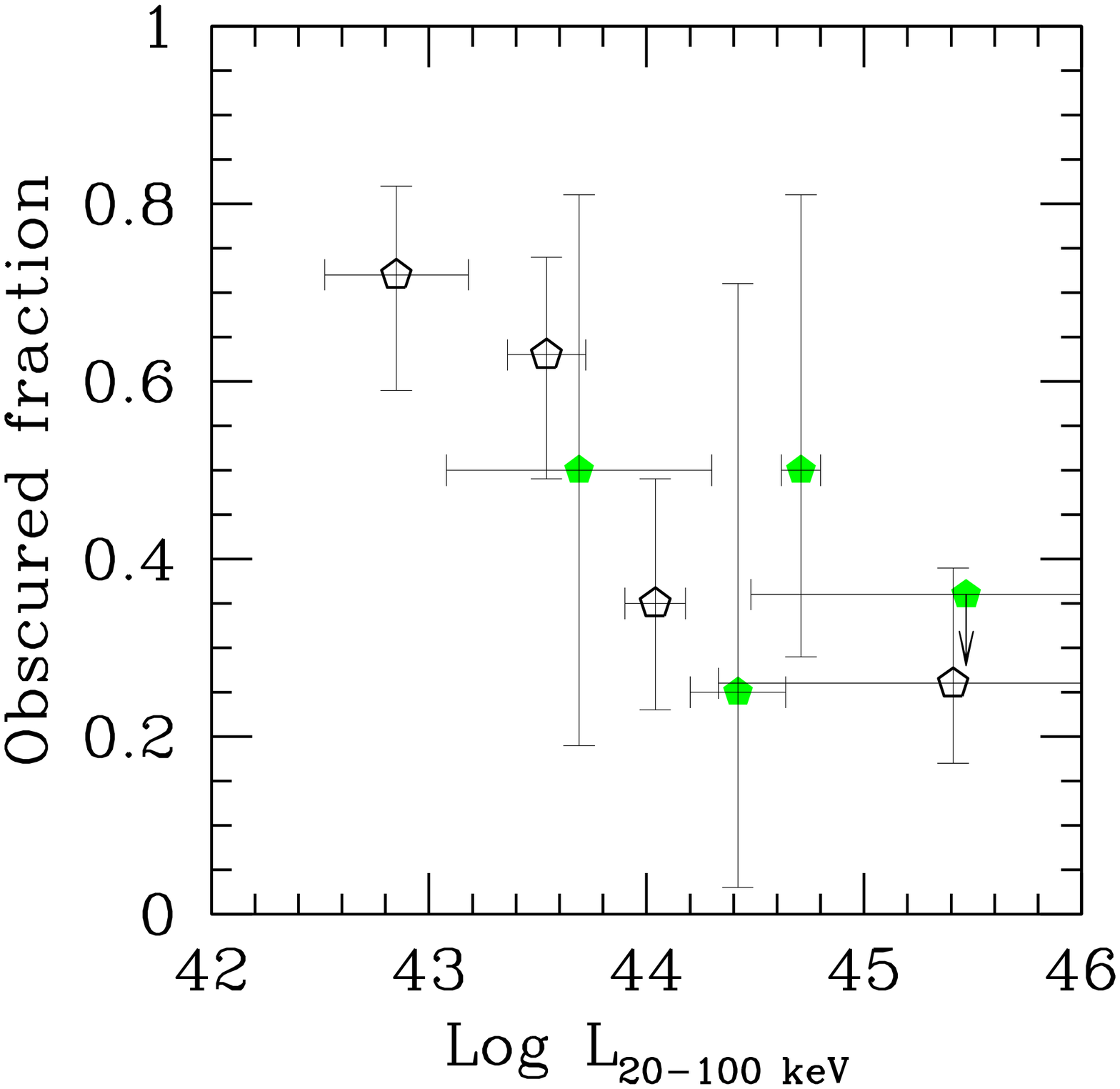}
}
\caption{Fraction of obscured AGN (number of objects with N$_{H}$ $>$ 10$^{22}$ cm$^{-2}$ over the total number) versus luminosity bins, 
computed for the 64 hard X-ray selected radio galaxies (left panel). Fraction of obscured AGN versus luminosity bins,
computed for the Malizia et al. (2012) sample (right panel). Non-radio galaxies objects are drawn as black empty polygons, while radio galaxies as green solid polygons.}
\label{obs}
\end{center}
\end{figure*}

We have explored a possible dependence of the observed fraction of absorbed AGN with luminosity.
In Figure~\ref{obs} we show the dependence of the absorbed fraction in our sample of 64 radio galaxies with the intrinsic hard X-ray luminosity (left panel).
Sources have been grouped in order to have nearly the same number of objects in each bin. 
We find a trend of decreasing obscured sources with luminosity, in agreement with the results from X-ray surveys. 
We have done the same exercise on the sample by Malizia et al. (2012), separating radio galaxies from non-radio galaxies (right panel). Unfortunately
the number of radio galaxies is very small, generating very large error bars, nonetheless, the fraction of
obscured radio galaxies seems to follow the same trend in luminosity as obscured non-radio galaxies. 
A dependence of obscured sources with redshift cannot be tested with our flux limited sample, as it would suffer from observational biases.
However, a possible evolution of the CT radio galaxies population with redshift was suggested by the work from 
Wilkes et al. (2013), who have estimated a fraction of CT AGN of 21 $\pm$ 0.7 \% in high redshift radio galaxies (1 $<$ z $<$ 2).
A confirmation to this result must come from the study of large samples at different redshifts ranges.

\section{Conclusions}

We have presented for the first time the observed column density distribution of a sample of hard X-ray selected radio galaxies.
The sample consists of 64 radio galaxies (mostly FRII), all within z$<$1, of relatively high luminosities (from 10$^{42}$ to 10$^{46}$ erg/s) and accreting at high Eddington ratios. 
All of them have an accurate measure of the column density and an accurate optical classification (except for two sources). 
Similarly to non-radio hard-X selected sources, we have found a fraction of $\sim$ 40\% of absorbed sources among the total sample,
and $\sim$ 75\% among type 2 AGN. On the other hand the observed fraction of CT AGN is $\sim$ 2-3\% compared to the 5-7\% found in non-radio galaxies hard X-ray selected AGN (Malizia et al. 2009, Burlon et al. 2011).
These results confirms that, at the zeroth order, unified models apply in a similar way to AGN with jets.
This is also confirmed by a recent study from Gupta et al. (2016) in which the dust covering factors of FRII and Quasars
show a very similar distribution, suggesting similar accretion conditions on parsec scales among the two classes.

Among radio galaxies, a handful of objects apparently deviate from the unified models predictions,
i.e., the optical classification do not match with the X-ray classification. Six objects display broad emission lines in their optical spectra
but their X-ray spectra are absorbed. However, the absorption usually comes from ionized material not associated with the classical torus, therefore the unified models predictions are not invalidated. One object is a type 2 AGN and has no X-ray absorption. Our simultaneous multi-frequency observations will allow to rule out variability as a possible
cause for such mismatch.

Finally, we have estimated the absorption and CT fractions in a sample of non-radio galaxies and radio galaxies extracted from the same sample selected at hard X-ray (from Malizia et al. 2012)
and therefore subject to the same biases effects. We find that the fraction of absorbed AGN is comparable among non-radio galaxies and radio galaxies (49$^{+6}_{-7}$\% and 35$^{+22}_{-17}$\%, respectively) 
while the fraction of CT AGN is only slightly larger in non-radio galaxies (6$\pm$3\%) compared to radio galaxies ($<$1. 4\%). We have selected one non-radio galaxies and one radio galaxies sub-sample in the same redshift range and then in the same flux range in order to exclude further biases, and overall we find that the fraction of absorbed AGN 
is always comparable (within errors) among the two groups, while there is a marginal evidence that 
the fraction of CT is always slightly smaller in radio galaxies. We have also checked for a possible dependence of the absorbed population on the luminosity  as suggested in the literature.
We found that the fraction of absorbed radio galaxies decreases with luminosity as well as non-radio galaxies. Unfortunately, the small number of sources tested here prevent us from drawing a firm conclusion. Ideally, statistically significant samples at different redshifts are needed to confirm such trend. 
INTEGRAL and \emph{Swift} are continuing in their survey of the hard X-ray sky reaching sensitivity levels down to 10$^{-12}$ ergs cm$^{-2}$ s$^{-1}$ allowing to detect new sources and to increase the sample of non-radio galaxies and radio galaxies (Bird et al. 2016, Mereminskiy et al. 2016, Malizia et al. 2016).
In the future, SKA surveys will trace the low-power radio-loud AGN population at a flux level of 1$\mu$Jy (Prandoni \& Seymour 2015) and in synergy with {\it Athena} 
it will be possible to investigate the evolution of the fraction of absorbed objects with X-ray luminosity and redshift up to z $\sim$ 4 - 5.
\\

{\bf Acknowledgements}\\

FP acknowledges support from INTEGRAL ASI/INAF n.2013-025-R.0. FP thanks Matteo Guainazzi and Paola Grandi for helpful discussions.
The authors acknowledge the use of NED, SIMBAD and HEASARC.

{}

\setcounter{table}{0}   
\begin{landscape}
\begin{table}
\scriptsize
\caption{\textbf{Radio Galaxies detected by \emph{INTEGRAL}/IBIS and \emph{Swift}/BAT}}
\label{tab1}
\begin{tabular}{lccccccclllll}
\hline
\textbf{Name} & \textbf{z} & \textbf{Opt Class} & \textbf{Radio} &  \textbf{Log $N_{\textrm H}$} & \textbf{F(2--10~{\textrm keV})}$^{\dagger}$  &
\textbf{F$_{\textrm {BAT}}$}$^{\dagger}$ &  \textbf{F$_{\textrm {IBIS}}$}$^{\dagger}$ & \textbf{Ref.} & Log M$_{BH}$ & \textbf{Ref.} & R$_{core}$ & \textbf{Ref.} \\
&&& \textbf{Morph} & & & (20--100~{\textrm keV})& (20--100~{\textrm keV}) &&&&\\
\hline
\hline
PKS 0018--19             &  0.095579      &  Sy1.9          &  FR II       &  21.89 [21.83 -- 21.95]  &  1.07   &  1.19   &   --      &  1   &  --     &  	--   & 0.186          & 1   \\
PKS 0101--649            &  0.163000      &  BLQSO          &  FR II       &  \textbf{20.36}          &  0.31   &  0.63   &   --      &  1   &  --     &  	--   & --		& --   \\
3C  033                  &  0.059700      &  Sy2            &  FR II       &  23.72 [23.69 -- 23.80]  &  0.27   &  1.71   &   --      &  2   &  8.48   &	1    & 0.003         & 2       \\
NGC 612                  &  0.029771      &  Sy2            &  FR I/II     &  24.04 [24.02 -- 24.06]  &  1.60   &  3.04   &   --      &  3   &  8.64   & 	1    & 0.009         & 3   \\
3C  059                  &  0.109720      &  Sy1.8          &  FR II       &  \textbf{20.78}          &  0.84   &  1.07   &   --      &  4   &  7.15   & 	2    & 0.034		& 4   \\
3C  062                  &  0.147000      &  Sy2            &  FR II       &  23.26 [23.18 -- 23.36]  &  0.2    &  0.77   &   --      &  5   &  8.81   & 	1    & 0.005         & 1       \\
4C +10.08                &  0.070000      &  NLRG           &  FR II       &  21.89[21.69 -- 22.05]   &  0.23   &  0.87   &   --      &  1   &  --     & 	--   & --		& --   \\
QSO B0309+411B           &  0.134000      &  Sy1            &  FR II       &  \textbf{21.11}          &  2.36   &  0.96   &  1.90     &  6   &  --     &   	--   & 3.0           &  5  \\
2MASX J03181899+6829322  &  0.090100      &  Sy1.9          &  FR II       &  22.61 [22.57 -- 22.66]  &  0.73   &  1.03   &  1.34     &  6   &  --     &   	--   & 0.133          &  6  \\
NGC 1275                 &  0.017559      &  Sy1.5          &  FR I        &  21.08 [21.04 -- 21.11]  &  1.23   &  3.80   &  3.00     &  6   &  8.5    &  	3    & 0.646          &  7  \\
3C 098                   &  0.0304        &  Sy2            &  FR II       &  23.08 [23.00 -- 23.18]  &  0.27   &   --    &  3.80     &  6   &  7.88   &  	4    & 0.0017$\pm$0.0005    & 8   \\
3C 105                   &  0.089000      &  Sy2            &  FR II       &  23.45 [23.40 -- 23.47]  &  0.21   &  1.69   &  3.10     &  6   &  8.61   & 	5    & 0.006          &	3\\
3C 109                   &  0.305600      &  Sy1.8          &  FR II       &  21.70 [21.60 -- 21.78]  &  0.80   &  0.82   &   --      &  7   &  8.30   & 	6    & 0.06$\pm$0.02       &    8 \\
3C 111                   &  0.048500      &  Sy1            &  FR II       &  21.66 [21.63 -- 21.69]  &  3.51   &  7.13   &  10.50    &  6   &  8.80   &   	7    & 0.04$\pm$0.02       &   8 \\
3C 120                   &  0.033010      &  Sy1            &  FR I?       &  21.10 [21.19 -- 21.22]  &  4.60   &  5.73   &  7.58     &  6   &  7.75	&   	7    & 0.67           &   3 \\
PKS 0442--28             &  0.147000      &  Sy2            &  FR II       &  21.93 [21.73 -- 22.07]  &  1.16   &  1.14   &   --      &  5   &  9.11	&  	5    & 0.041          & 9	\\
Pic A                    &  0.035058      &  Liner/Sy1      &  FR II       &  \textbf{20.78}          &  1.13   &  2.36   &  3.24     &  6   &  8.70	&  	7    & 0.07           & 3    \\
PKS 0521--364            &  0.056546      &  Sy1            &  FR I/II     &  \textbf{20.55}          &  1.10   &  2.00   &  2.19     &  6   &  8.6	&   	8    & 0.176         &  3  \\
PKS 0707--35             &  0.110800      &  Sy2            &  FR II       &  22.71[22.38 -- 22.72]   &  0.56   &  1.19   &   --      &  8   &  --     & 	--   & --		&--    \\    
3C 184.1                 &  0.118200      &  Sy2            &  FR II       &  23.03[22.88 -- 23.16]   &  0.26   &  0.69   &   --      &  9  &  8.32    & 	9    & --         &  --  \\
B3 0749+460A             &  0.051799      &  Sy1.9          &  FR II       &  \textbf{20.25}          &  0.30   &  0.84   &   --      &  1   &  6.9     &	10   & --		&--    \\   
3C 206                   &  0.197870      &  Sy1.2          &  FR II       &  \textbf{20.75}          &  1.50   &  1.52   &   --      &  8   &  8.86    & 	11   & 0.273          &   1  \\ 
4C +29.30                &  0.064715      &  Sy2            &  FR I/II     &  23.70 [23.68 -- 23.71]  &  0.11   &  0.95   &   --      &  10  &  8.2     & 	12   & 0.429$\pm$0.134      & 8   \\
3C 227                   &  0.086272      &  Sy1.5          &  FR II       &  22.16 [22.10 -- 22.22]  &  2.36   &  1.52   &   --      &  5   &  8.90	&	7    & 0.0123         &	3\\  
VII Zw 292               &  0.058100      &  Sy2            &  FR II       &  23.96 [23.80 -- 24.16]  &  0.08   &  0.94   &   --      &  11  &  --     & 	--   & 0.006          &  10  \\
3C 234                   &  0.184925      &  Sy1.9          &  FR II       &  23.81 [23.79 -- 23.83]  &  0.13   &  0.48   &   --      &  12  &  8.99    &   	13   & 0.02$\pm$0.006      &  8  \\
PKS 1143--696            &  0.244000      &  Sey1.2         &  FR II       &  \textbf{21.21}          &  0.54   &  0.92   &  1.28     &  6   &  8.62    & 	14   & --		&  --  \\
IGRJ13107--5626          &  -             &  -              &  FR II?      &  23.59 [23.41 -- 23.80]  &  0.11   &  1.02   &  0.87     &  13  &  --     & 	--   & --		&   -- \\    
CenA                     &  0.00088       &  Sey2           &  FR I        &  23.17 [23.16 -- 23.18]  &  21.20  &  82.73  &  59.00    &  6   &  9.11   & 	15   & 0.125          &   3 \\
CenB                     &  0.012916      &  NLRG           &  FR I/II     &  21.11 [21.98 -- 22.21]  &  0.49   &   --    &  1.77     &  6   &  --     &   	--   & 0.07$\pm$0.02          &   8  \\
3C 287.1                 &  0.2156        &  Sy 1           &  FR II       &  \textbf{21.21}          &  0.28   &   --    &  1.50     &  6   &  8.49   &   	16   & 0.365$\pm$0.114      &   8 \\
2MASX J14364961--1613410 &  0.144537      &  BLQSO          &  FR I/II     &  \textbf{20.88}          &  0.72   &  1.08   &   --      &  1   &  8.64   & 	17   & --              &   -- \\
IGR 14488--4008          &  0.123         &  Sey1.2         &  FR II       &  22.89 [22.74 -- 23.04]  &  0.53   &  0.82   &  0.60     &  14  &  8.58   & 	14   & --              & --   \\
3C 309.1                 &  0.905000      &  Sy1.5          &  C           &  $<20.50$                &  0.17   &  0.54   &   --      &  15  &  9.1     & 	18   & 1.87$\pm$0.58       & 8     \\
4C +63.22                &  0.20400       &  Sy1            &  FR II       &  \textbf{20.17}          &  0.36   &  0.56   &   --      &  1   &  --     &  	--   & 0.028          &	 10 \\
3C 323.1                 &  0.264300      &  Sy1.2          &  FR II       &  20.78 [20.00 -- 21.04]  &  0.55   &  0.64   &   --      &  16  &  9.12     & 	19   & 0.046$\pm$0.014      &   8 \\
4C +23.42                &  0.118000      &  Sy1            &  FR I        &  21.75 [21.60 -- 21.86]  &  0.32   &  0.79   &   --      &  1   &  --     &  	--   & --		&  --  \\
Leda 100168              &  0.183000      &  Sy1            &  FR II       &   $<20.69$               &  0.43   &  0.74   &   --      &  1   &  --     &  	--   & --		& --   \\
3C 332                   &  0.151019      &  Sy1            &  FR II       &   \textbf{20.30}         &  0.58   &  1.34   &   --      &  4   &  8.64	& 	16   & --         & --   \\
Mrk 1498                 &  0.054700      &  Sy1.9          &  FR II       &   23.76 [23.67 -- 23.84] &  0.90   &  2.69   &   --      &  17  &  8.59     & 	20   & 0.09           &  11  \\    
4C +34.47                &  0.206000      &  Sy1            &  FR II       &   \textbf{20.49}         &  0.94   &  1.24   &   --      &  18  &  8.01	& 	11   & 0.86$\pm$0.26       &  8  \\
PKS 1737--60             &  0.410000 Phot &  -              &  FR II       &   \textbf{20.77}         &  0.73   &  0.84   &   --      &  1   &  --     & 	--   & --	        & --   \\
4C +18.51                &  0.186000      &  Sy1            &  FR II       &   \textbf{20.74}         &  0.49   &  0.59   &   --      &  1   &  9.35	&   	21   & 0.1            &  12  \\
IGR J17488--2338         &  0.240         &  Sy1.5          &  FR II       &   22.06 [21.96 -- 22.15] &  0.20   &   --    & 1.40      &  19  &  9.11    & 	14   & 0.179          &  13   \\  
\hline																				       
\hline																				       
\end{tabular}																			       
\end{table}																			       
\end{landscape} 																		       
																				       
\begin{landscape}																		       
\begin{table}																			       
\scriptsize																			       
\begin{tabular}{lccccccclllll}																	       
\hline																				       
\textbf{Name} & \textbf{z} & \textbf{Opt Class} & \textbf{Radio} &  \textbf{Log $N_{\textrm H}$} & \textbf{F(2--10~{\textrm keV})}$^{\dagger}$  &
\textbf{F$_{\textrm {BAT}}$}$^{\dagger}$ &  \textbf{F$_{\textrm {IBIS}}$}$^{\dagger}$ & \textbf{Ref.} & Log M$_{BH}$ & \textbf{Ref.} & R$_{core}$ & \textbf{Ref.} \\
&&& \textbf{Morph} & & & (20--100~{\textrm keV})& (20--100~{\textrm keV}) &&&&\\
\hline																				       
\hline
3C 380                   &  0.692000      &  Sy1.5          &  FR II       &  \textbf{20.75}          &  0.40   &  0.96   &  --   &  20 &  9.4  	& 	18  &       0.817$\pm$0.254   & 8   \\
3C 382                   &  0.057870      &  Sy1            &  FR II       &  \textbf{20.79}          &  5.98   &  5.32   & 4.24  &  6  &  8.90  	&    	21  &       0.05$\pm$0.02    & 8  \\
3C 390.3                 &  0.056100      &  Sy1.5          &  FR II       &  \textbf{20.63}          &  2.14   &  6.20   & 5.90  &  6  &  8.80  	&    	21  &       0.09$\pm$0.03    & 8  \\
PKS 1916--300            &  0.166819      &  Sy1.5/1.8      &  FR II       &  \textbf{20.90}          &  0.62   &  0.97   & 0.99  &  6  &  --  		&    	--  &       0.02   	  &  14 \\
3C 403                   &  0.059000      &  Sy2            &  FR II       &  23.65 [23.58 -- 23.62]  &  1.30   &  1.96   & 1.87  &  6  &  8.96 	& 	5   & 	    0.004  	  & 3	 \\
Cygnus A                 &  0.056075      &  Sy1.9          &  FR II       &  23.30 [23.25 -- 23.32]  &  1.20   &  8.29   & 8.49  &  6  &  9.40 	&    	22  &       0.002  	  & 3  \\
PKS 2014--55             &  0.060629      &  Sy2            &  FR I        &  23.51 [23.36 -- 23.64]  &  0.39   &  1.94   &  --   &  1  &  --  		&  	--  &       --      &  --    \\
4C +21.55                &  0.173500      &  Sy1            &  FR II       &  21.78 [21.62 -- 22.09]  &  1.40   &  1.98   & 2.91  &  21  &  --  	&   	--  &       --      & --  \\
4C +74.26                &  0.104000      &  Sy1            &  FR II       &  21.15 [21.04 -- 21.20]  &  2.53   &  3.18   & 4.16  &  6  &  9.37  	&    	7   &       0.95$\pm$0.29    &  8 \\
S5 2116+81               &  0.084000      &  Sy1            &  FR I        &  \textbf{21.38}          &  1.21   &  2.14   & 3.04  &  6  &  8.12  	&    	7   &       1.80$\pm$0.54    &  8 \\
4C 50.55                 &  0.020000      &  Sy1            &  FR II       &  22.89 [22.79 -- 22.99]  &  4.88   &  12.25  & 12.50 &  6  &  7.80 	& 	23  & 	    0.1    	   & 15	    \\
3C 433                   &  0.101600      &  Sy2            &  FR I/FR II  &  22.94 [22.90 -- 23.97]  &  0.10   &  1.02   &  --   &  22 &  9.10 	&  	24  &       0.0011$\pm$0.0003  & 8  \\
PKS 2135--14             &  0.20047       &  Sy1.5          &  FR II       &  23.26 [23.02 -- 23.48]  &  0.57   &  0.90   &  --   &  5  &  9.65  	&  	5   &       0.096    	   & 3  \\
PKS 2153--69             &  0.028273      &  Sy1            &  FR II       &  \textbf{20.40}          &  0.72   &  0.84   &  --   &  9 &  --  		& 	--  & 	    0.03$\pm$0.01    & 	 8   	 \\
2MASX J22194971+2613277  &  0.085000      &  Sy1            &  FR II       &  21.40 [21.30 -- 21.49]  &  0.67   &  1.03   &  --   &  1  &  --  		&  	--  &       --      &  -- \\
3C 445                   &  0.055879      &  Sy1.5          &  FR II       &  23.35 [23.26 -- 23.46]  &  0.67   &  2.63   &  --   &  23 &  8.89  	& 	5   & 	    0.04   	   &   3  \\  
3C 452                   &  0.081100      &  Sy2            &  FR II       &  23.77 [23.70 -- 23.83]  &  0.50   &  1.97   &  2.75 &  6  &  8.54  	& 	13  &       0.06$\pm$0.018   &   8\\ 
PKS 2300--18             &  0.128929      &  Sy1            &  FR II?      &  \textbf{20.34}          &  0.51   &  0.70   &  --   &  1  &  8.25  	& 	25  & 	    1.07   	  &  1 \\
PKS 2331--240            &  0.047700      &  Sy2            &  FR II       &  \textbf{20.21}          &  0.71   &  0.85   &  --   &  1  &  --  		& 	--  & 	    0.65   	  &  16  \\
PKS 2356-61              &  0.096306      &  Sy2            &  FR II       &  23.17 [23.13 -- 23.21]  &  0.20   &  0.91   &  --   &  5  &  8.96  	&  	5   &       0.008  	  & 3  \\ 
\hline
\hline
\end{tabular}
\begin{list}{}{}
\item $^{\dagger}$ The 2--10 and 20--100 keV fluxes are in unit of $10^{-11}$ erg cm$^{-2}$ s$^{-1}$. X-ray references: 1. This work; 2. Torresi et al. (2009);
3. Eguchi et al. (2011); 4. Tombesi et al. (2014); 5. Mingo et al. (2014); 6. Malizia et al. (2012); 7. Hardcastle et al (2006); 8. Tazaki et al. (2013); 9. Grandi et al. (2006);
10. Sobolewska et al. (2012); 11. Evans et al. (2008); 12. Vasudevan et al. (2013); 13. Landi et al. (2010); 14. Molina et al. (2015); 15. Siemiginowska et al. (2008); 16. Hasenkopf et al. (2002);
17. Eguchi et al. (2009); 18. Page et al. (2004); 19. Molina et al. (2014); 20. Belsole et al. (2006);  21. Malizia et al. (2016); 22. Miller \& Brandt (2009); 23. Sambruna et al. (2007). 
Black hole mass references: 1. Bettoni et al. (2003); 2.Wandel (2002); 3.Wilman et al.(2005); 4. Woo \& Urry (2002); 5. Mingo et al.(2014); 6. McLure et al. (2006); 7. Kataoka et al. (2011); 
8. Sbarrato et al.(2012); 9. Grandi et al. (2006); 10. Greene \& Ho (2007); 11. Liu et al. (2006); 12. Sobolewska et al. (2012); 13. Marchesini et al. (2004); 14. Masetti et al.(2013);  
15. Sadoun \& Coulin (2012); 16. Mezcua et al. (2011); 17. Kim et al. (2008); 18. Wu (2009); 19. Runnoe et al. (2013);  20. Winter et al. (2010); 21. Wang, Mao \& Wei (2009); 22. Tadhunter et al. (2003);
23. Tazaki et al. (2010); 24. Cao \& Rawlings (2004); 25. Wu \& Liu (2004). 5 GHz radio core dominance references: 1. Reid et al. (1999); 2. Hardcastle et al. (1998); 3. Morganti et al. (1993);
4. Bondi et al. (1993); 5. De Bruyn et al. (1989); 6. Schoenmakers et al. (1998); 7. Abdo et al. (2010); 8. Fan \& Zhang (2003); 9.  Morganti et al. (1999); 10. Lara et al. (2001), at 1.4 GHz; 
11. Rottgering et al. (1996); 12. Lister, Gower \& Hutchings (1994); 13. Molina et al. (2014); 14. Duncan \& Sprooats (1992); 15. Molina et al. (2007); 16. Slee et al (1994).
\end{list}
\end{table}                         
\end{landscape}

\appendix

\section{\emph{Swift}/XRT data reduction and analysis}

We took all radio galaxies from the sample of Bassani et al. (2016) and extracted 13 objects for which X-ray spectral information were not found in the literature. 
This list of sources was then cross-correlated with the \emph{Swift}/XRT archival dataset, and for all sources 
we found X-ray observations available in the archive. 
The log of all X-ray observations analysed in this work is given in Table~\ref{tab2a}, where we report for each individual radio galaxy, the XRT
observation ID, the date and the exposure of the XRT pointings.

The XRT data were processed using the XRTDAS standard data pipeline package (\textsc{xrtpipeline} v. 0.12.9) 
to produce calibrated and cleaned PC-mode event file. Observations were summed together using 
\textsc{XSELECT v. 2.4c}.
XRT images in the 0.3--10 keV energy band were obtained and analysed by means of the software package 
\textsc{XIMAGE} v. 4.5.1.
We found that all sources were detected by XRT, although not all with good statistical significance.

For the spectral analysis, source events were extracted within a circular region with a radius of 20 
pixels (1 pixel $\sim$ 2.36 arcsec) centred on the source position, while background events were extracted 
from a source-free region close to the X-ray source of interest. The spectra were obtained from the 
corresponding event files using the \textsc{XSELECT v. 2.4c} software and binned using \textsc{grppha} in 
an appropriate way, so that the $\chi^{2}$ statistic could be applied. We used version v.014 of the 
response matrices and created individual ancillary response files \emph{arf} using \textsc{xrtmkarf v. 
0.6.0}.

Next, we combined the XRT spectra with the BAT ones (from the 70-month BAT catalogue, Baumgartner et al. 
2013)\footnote{available at: http://swift.gsfc.nasa.gov/results/bs70mon/.} to perform the spectral 
analysis over the 0.3--100 keV energy range. We adopted a simple absorbed power law model to fit the 
source spectra where absorption is partly due to a Galactic column density in the source direction (fixed 
value) and partly to an intrinsic column density (free parameter). We also introduced in the fitting 
procedure a cross-calibration constant ($C_{\it calib}$) to account for a possible mismatch between XRT 
and BAT data, as well as for source flux variations.

The results of the spectral fittings are shown in Table~\ref{tab3a} where we report for each source
the value of the Galactic column density,  
the photon index, the 2--10 keV Luminosity, the $\chi^{2}/\nu$ values of the fit, and the 
XRT/BAT cross-correlation constant, while 
intrinsic absorption values, as well as 2--10 keV fluxes are given in Table 1 of the main text. 
In the fitting procedure, when no absorption is measured, the 
Galactic column density value is used as an upper limit to the amount of absorption.

Those sources for which this simple model does not fit the data properly, indicating the need for 
an extra feature, are briefly discussed in the following sections.

\textbf{4C +63.22}\\
For this source, our baseline model does not provide a good fit to the data ($\chi^{2}/\nu = 69.0/43$) as 
residuals, indicative of an absorbing feature at around 0.9 keV, are clearly visible in 
the data-to-model ratio. We therefore tested our data for the presence of an absorption 
edge, leaving all the parameters free to vary: the fit improves significantly ($\chi^{2}/\nu = 38.4/41$) 
and the component is required by the data at more than 99.99\% confidence level (c.l.). The
absorption edge turned out to be at $0.94\pm0.06$ and have $\tau_{max} = 0.91^{+0.34}_{-0.30}$, thus 
indicating an origin from Fe \textsc{XX}. This 
result is not surprising as Molina et al. (2015) have recently detected absorption features from ionised 
elements in the giant radio galaxy IGR J14488--4008.

\textbf{PKS 2014$-$55}\\
This is another case in which the baseline model do not provide a good fit to the data ($\chi^{2}/\nu = 
35.8/12$) as an excess below 1 keV is observed in the residuals. Also in this case we added to the baseline model
a second power law passing through intrinsic absorption, having the
same photon index of the primary one, which yields an improvement ($>$ 99.99\%) of the fit as shown in Table~\ref{tab3a}

\textbf{PKS 2300$-$18}\\
For this source, there is a hint of an excess around 6 keV, again due to neutral iron, which we modelled by adding a narrow 
line component. This feature is required at around 99.8\% ($\Delta\chi^{2}/\nu = 14.6/2$) and provides a line centroid
at $6.51\pm0.10$ and an $EW = 608^{+267}_{-272}$ eV.

\begin{table*}
\begin{minipage}{155mm}
\centering
\scriptsize
\caption{Log of the \emph{Swift}/XRT observations used in this paper.}
\label{tab2a}
\begin{tabular}{lccc}
\hline
\hline
Source name                &     ID        &   Obs date     & Exposure       \\
                           &               &                &      (s)        \\
\hline
\hline
PKS 0018$-$19              &  00040886001  &  Sep 27, 2010  &  8037  \\
                           &  00040691001  &  Jan 30, 2011  &  6494  \\
total obs                  &      --       &       --       &  14531  \\
\hline                  
PKS 0101$-$649             &  00047109001  &  Nov 01, 2011  &  864   \\
                           &  00047109002  &  Nov 11, 2011  &  2778  \\
                           &  00047109003  &  Nov 17, 2011  &  166   \\
                           &  00047109004  &  Nov 20, 2011  &  241   \\
                           &  00047109005  &  Nov 23, 2011  &  637   \\
                           &  00047109006  &  Nov 25, 2011  &  2641  \\
                           &  00047109007  &  Apr 05, 2012  &  2101  \\
total obs                  &      --       &       --       &  9428  \\
\hline                
4C $+$10.08                &  00037347001  &  Dec 05, 2010  &  527    \\
                           &  00037347002  &  Dec 14, 2010  &  6956   \\
                           &  00037347003  &  Mar 02, 2011  &  1143   \\
                           &  00037347004  &  Mar 06, 2011  &  2061   \\
total obs                  &               &                &  10687  \\
\hline                   
B3 0749$+$460A             &  00037357001  &  Feb 24, 2008  &  9353   \\
\hline
2MASX J14364961$-$1613410  &  00040978001  &  Dec 25, 2010  &  595    \\
                           &  00040978002  &  Apr 12, 2011  &  222   \\
                           &  00047122001  &  Apr 12, 2012  &  193  \\
                           &  00047122003  &  May 08, 2013  &  632    \\ 
                           &  00047122004  &  May 10, 2013  &  110    \\
                           &  00047122005  &  May 20, 2013  &  865    \\
                           &  00047122006  &  May 21, 2013  &  2775    \\
                           &  00047122007  &  May 26, 2013  &  341    \\
                           &  00047122008  &  May 27, 2013  &  3874    \\
total obs                  &               &                &  9607    \\
\hline                  
4C $+$63.22                &  00035401002  &  Oct 17, 2006  &  5546   \\
\hline
4C $+$23.42                &  00038073001  &  Sep 25, 2008  &  2269    \\
                           &  00038073002  &  Sep 26, 2008  &  4930    \\
                           &  00038073004  &  Jan 04, 2009  &  6690    \\  
total obs                  &               &                &  13889   \\
\hline
LEDA 100168                &  00037998001  &  Jul 05, 2008  &  4885   \\
                           &  00037998002  &  Jul 06, 2008  & 2563   \\
                                                      &  00037998002  &  Jul 06, 2008  &  2563   \\
                           &  00037998002  &  Jul 07, 2008  &  2532  \\
total obs                  &               &                &  9980   \\
\hline
PKS 1737$-$60              &  00047123001  &  Nov 10, 2011  &  836    \\
                           &  00047123002  &  Nov 11, 2011  &  5204   \\
                           &  00047123003  &  Nov 14, 2011  &  1123   \\
                           &  00047123004  &  Nov 17, 2011  &  2871   \\
total obs                  &               &                &  10034   \\
\hline
4C $+$18.51                &  00041782001  &  Oct 31, 2010  &  2490   \\
                           &  00041782002  &  Jan 13, 2011  &  1586   \\
                           &  00041782003  &  Jan 21, 2011  &  1715   \\
                           &  00041782004  &  Jan 25, 2011  &  1680   \\
                           &  00041782004  &  Jan 28, 2011  &  3003  \\
total obs                  &               &                &  10474  \\
\hline
2MASX J22194971$+$2613277  &  00041122002  &  Jul 12, 2010  &  3181  \\
                           &  00041122003  &  Jul 13, 2010  &  7165  \\ 
                           &  00041122004  &  Jul 15, 2010  &   946  \\ 
                           &  00041122005  &  Jul 17, 2010  &  1991   \\
total obs                  &               &                &  13283  \\
\hline
PKS 2331$-$240             &  00031731001  &  Jun 05, 2010  &  1668  \\
                           &  00031731002  &  Jun 05, 2010  &  3924  \\   
                           &  00041128001  &  Sep 29, 2010  &  9052  \\
total obs                  &               &                &  14644  \\
\hline
\hline
\end{tabular} 
\end{minipage}
\end{table*}

\begin{table*}
\begin{minipage}{155mm}
\centering
\caption{XRT spectral analysis results.}
\label{tab3a}
\scriptsize
\begin{tabular}{lcccccc}
\hline
\hline
Source name                 &  N$_{\textrm H(Gal)}^{a}$ & $\Gamma$       &  L(2--10 keV)$^{b}$ & $\chi^{2}/\nu$  & $C_{\textrm calib}^{c}$  &  Model \\
\hline
\hline
PKS 0018$-$19                 &   0.0201         & $1.72\pm0.12$  &  2.37       &  105.5/109      & $0.55^{+0.22}_{-0.16}$ & 
absorbed pl  \\
\hline
PKS 0101$-$649                &   0.0231         & $1.69\pm0.09$  &  2.16      &  33.8/48        & $1.00^{+0.47}_{-0.34}$ & 
simple pl  \\
\hline
4C $+$10.08                   &   0.100          & $1.70^{+0.31}_{-0.29}$  &  0.262    &  17.7/21  & $1.78^{+2.00}_{-0.99}$ &
absorbed pl  \\
\hline
B3 0749$+$460A                &   0.0179         & $1.64\pm0.08$           &  0.188    &  61.9/49  & $1.22^{+0.48}_{-0.36}$  &
simple pl  \\
\hline
2MASX J14364961$-$1613410     &   0.0759         & $1.72\pm0.06$           &  3.90     &  74.5/93  & $0.79^{+0.29}_{-0.23}$  &
simple pl  \\
\hline
4C $+$63.22                   &   0.0148         & $1.96\pm0.08$           &  4.30     &  38.4/41  & $1.38^{+0.63}_{-0.46}$  &
simple pl + absorption edge  \\
\hline
4C $+$23.42                   &   0.0416         & $1.67\pm0.21$           &  1.11     &  46.3/38  & $1.07^{+0.91}_{-0.50}$  &
absorbed pl   \\
\hline
LEDA 100168                   &   0.0593         & $1.75\pm0.11$           &  3.91     &  67.4/65  & $0.97^{+0.45}_{-0.31}$  &
absorbed pl   \\        
\hline
PKS 1737$-$60                 &   0.0584         & $1.94\pm0.07$           &  23.3     &  86.1/75  & $1.55^{+0.72}_{-0.56}$  &
simple pl  \\
\hline
4C $+$18.51                   &   0.0551         & $1.83\pm0.07$           &  4.72     &  66.3/75  & $0.81^{+0.41}_{-0.33}$  &
simple pl  \\
\hline
PKS 2014$-$55                 &   0.0481         & $1.86\pm0.21$           &  0.304    &  17.2/10  & $1.01^{+0.72}_{-0.42}$  &
double pl   \\
\hline
2MASX J22194971$+$2613277     &   0.0664         & $1.69\pm0.11$           &  1.18     &  70.1/85   & $0.73^{+0.31}_{-0.22}$  &
abserbed pl   \\
\hline
PKS 2300$-$18                 &   0.0217         & $1.83\pm0.21$           &  2.24     &  139.0/117 & $1.01^{+0.39}_{-0.33}$  &
simple pl + iron line \\
\hline
PKS 2331$-$240                &   0.0163         & $1.70\pm0.04$           &  0.375    &  147.5/138 & $0.60^{+0.18}_{-0.16}$  &
simple pl   \\
\hline
\hline
\end{tabular}
\begin{list}{}{}
\item $^{a}$ In units of $10^{22}$ cm$^{-2}$ (from Kalberla et al. 2005);
\item $^{c}$ IBIS/XRT cross-calibration constant;
\item $^{b}$ In units of $10^{44}$ erg s$^{-1}$.
\end{list}
\end{minipage}
\end{table*}

\end{document}